\documentclass[prl,twocolumn,tightenlines,amsmath,showpacs]{revtex4}
\usepackage{graphicx}

\newcommand{\beq}{\begin{equation}}
\newcommand{\eeq}{\end{equation}}
\newcommand{\bea}{\begin{eqnarray}}
\newcommand{\eea}{\end{eqnarray}}
\newcommand{\bml}{\begin{subequations}}
\newcommand{\eml}{\end{subequations}}
\newcommand{\ba}{\begin{array}}
\newcommand{\ea}{\end{array}}
\newcommand{\bc}{\begin{center}}
\newcommand{\ec}{\end{center}}
\newcommand{\commentout}[1]{{}}
\newcommand{\bk}{{\bf k}}

\newcommand{\half}{\hbox{$\frac{1}{2}$}}
\newcommand{\quarter}{\hbox{$\frac{1}{4}$}}

\newcommand{\etal} {{\it et al.\/}}

\newcommand{\vol}[1]{{\bf #1}}
\newcommand{\comment}[1]{{}}

\begin{document}


\title{Many-Body Rate Limit on Photoassociation of a Bose-Einstein Condensate}
\author{Matt Mackie and Pierre Phou}
\affiliation{Department of Physics, Temple University, Philadelphia, PA 19122}
\date{\today}

\begin{abstract}

We briefly report on zero-temperature photoassociation of a Bose-Einstein condensate, focusing on the many-body rate limit for atom-molecule conversion. An upgraded model that explicitly includes spontaneous radiative decay leads to an unanticipated shift in the position of the photoassociation resonance, which affects whether the rate (constant) maximizes or saturates, as well as the limiting value itself. A simple analytical model agrees with numerical experiments, but only for high density. Finally, an explicit comparison with the two-body unitary limit, set by the size of the condensate, finds that the many-body rate limit is generally more strict.

\end{abstract}

\maketitle

{\em Introduction.}--Quantum degenerate molecules are of interest for investigations fundamental constants~\cite{HUD06} and physical chemistry~\cite{SKO99}, quantum computing and encryption applications~\cite{DEM02}, as well as condensed-matter~\cite{BUE07} and quantum-gravity analogues~\cite{NAT10}. For alkali-metal atoms~\cite{COR02}, quantum degeneracy occurs around 100~nK, and is achieved by laser and evaporative cooling atoms that begin at an oven temperature of about 600~K. Unfortunately, laser cooling depends on the existence of a {\em closed} few-level system, which does not exist for molecules, making quantum degeneracy harder to achieve with laser cooling~\cite{SHU09}. Coherent association--via either laser or magnetic fields--can provide an end run around the need for molecular laser cooling: Since it is efficient at quantum degenerate phase space densities~\cite{BUR96}, and since it is fundamentally coherent, association creates molecular condensate from already-condensed atoms~\cite{JAV99}. Whereas magnetoassociation is currently in vogue~\cite{NI08}, photoassociation is crucial for systems that do not possess a Feshbach resonance, such as alkali-earth atoms.

The idea of coherent photoassociation raises a fundamental question of its own, namely the rate limit for converting atoms into molecules. Simply put, photoassociation depletes the pair wavefunction, and the rate limit on atom-molecule conversion is set by the refill time. In a many-body model~\cite{JAV02}, the refill time is determined by the interparticle spacing, and the maximum rate per unit density is approximately $K=R/\rho\sim\hbar\rho^{-1/3}/m_r$, where $\rho$ is the condensate density and $m_r$ is the reduced mass of a dissociated pair of atoms. In a two-body model~\cite{BOH99}, the de~Broglie wavelength sets the refill time, so that $K=R/\rho\sim\hbar\Lambda_D/m_r$, which is also known as the unitary limit and, at zero-temperature, is set by the size of the condensate. The two limits should merge at the onset of quantum degeneracy, since $\rho\Lambda_D^3\sim1$, but the deeply degenerate regime has not been fully investigated. Also, both the many-body and unitary rates reach a maximum for strong atom-molecule coupling. However, the many-body model does not include spontaneous radiative decay, and this qualitative agreement should be confirmed.

Experiments above~\cite{SCH02} or at~\cite{PRO03} the quantum degenerate limit are inherently inconclusive, as mentioned above, whereas the first experiments with a condensate were thwarted by strong dipole forces at strong laser coupling~\cite{MCK02}. Subsequent condensate experiments combined photoassociation and the Feshbach resonance, which enhances photoassociation and thereby allowed for observation of the rate limit at manageable intensities~\cite{JUN08}, but observations were again consistent with both the rogue and unitary model~\cite{MAC08}. Recent theory has shown that, up to an interference factor, the many-body result for Feshbach-photoassociation agrees reasonably with the result for photoassociation alone, but only agrees with the unitary limit in a small magnetic-field window near the Feshbach resonance, and is otherwise more strict~\cite{MAC10}.

Here we briefly report on the many-body rate limit for photoassociation of a Bose-Einstein condensate at zero temperature, based on an upgrade to our previous model~\cite{JAV02} to include spontaneous radiative decay. Numerical experiments reveal an unanticipated lightshift in the position of laser resonance, and the effect of this lightshift on the photoassociation rate is examined for high and low condensate densities. We use LiNa as an example, but the results apply generally to other species, including homonuclear pairs, as well as the Feshbach resonance.

\begin{figure}[b]
\centering
\includegraphics[width=8.75cm]{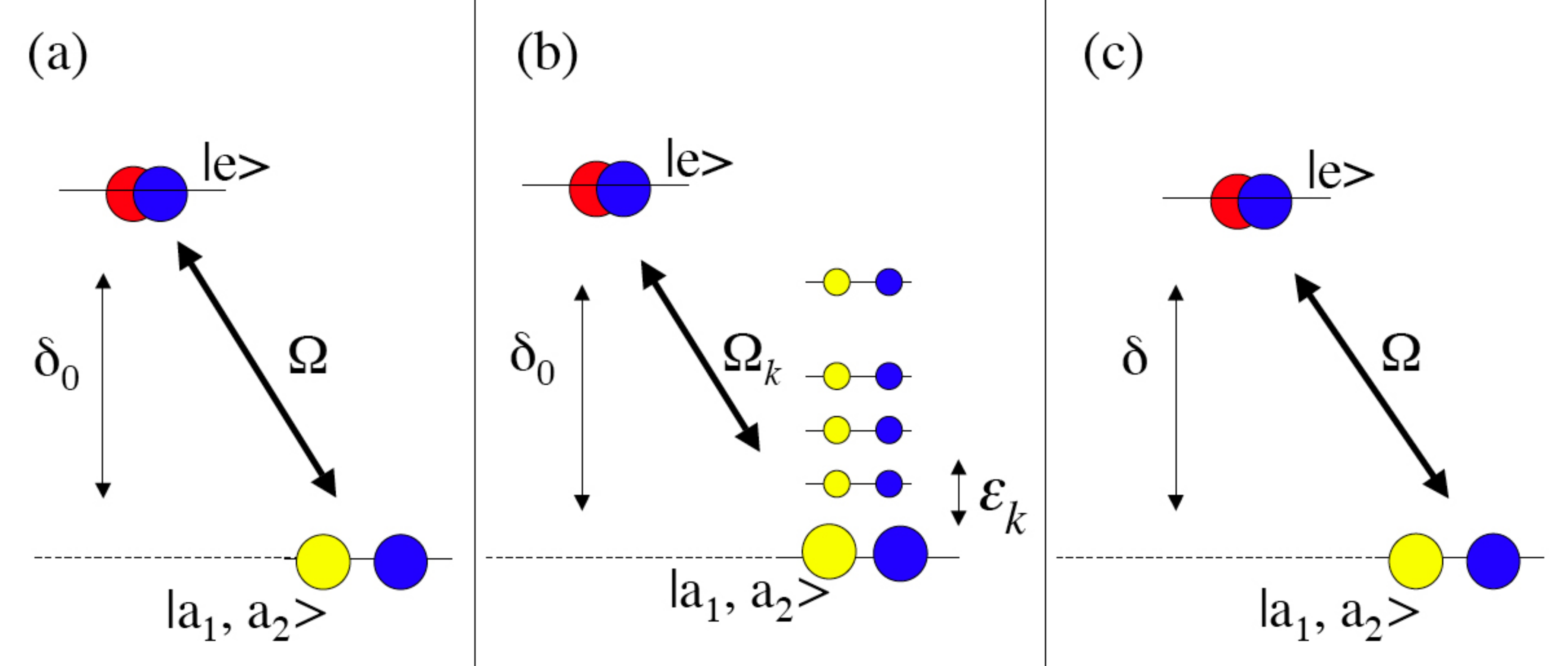}
\caption{(Color online) Few-level illustration for heteronuclear photoassociation. (a) Basic free-bound transition from the joint condensate to the electronically-excited molecular state, where $\delta_0$ is the laser detuning from resonance and $\Omega$ is the atom-molecule coupling. (b) Quasicontinuum levels account for photodissociation to atom pairs with equal and opposite momentum, where $\Omega_\bk=\Omega f_\bk$. (c) For steady-state photodissociation, the system is effectively a two-level system with a lightshifted detuning $\delta$.}
\label{FEWL}
\end{figure}

{\em Model}--We focus on the many-body model for heteronuclear photoassociation~\cite{SHA99} of a Bose-Einstein condensate of two miscible~\cite{TIM97} species of atoms. Photoassociation generally occurs on a timescale much shorter than the dipole-dipole and trap interactions, and these are therefore negligible. Consider $N_1$ atoms of mass $m_1$ and $N_2$ atoms of mass $m_2$ that have each Bose-condensed into the zero-momentum state $|a_1,a_2\rangle=|0_1\rangle|0_2\rangle$, with the total number of atoms $N=N_1+N_2$. The photoassociation laser then destroys an atom from each species and creates a dipolar molecule of mass $m_3=m_1+m_2$ in the electronically-excited state $|e\rangle$, with the laser detuning $\delta>0$ indicating an open dissociation channel, as per Fig.~\ref{FEWL}(a). The condensates are assumedly miscible, so that heteronuclear photoassociation can proceed. In the language of second quantization,  annihilation of an atom (molecule) from the $i$-th atomic (molecular) condensate is represented by the operator $a_{i,0}\equiv a_i$ ($b_0\equiv b$). The simplest theory has only these three levels/modes, and dissociation of the molecules can only take place back into the original atomic condensate level. To be more precise, molecular dissociation to noncondensate levels should be included, as per Fig.~\ref{FEWL}(b). This situation arises because a condensate molecule need not dissociate back to the atomic condensate, but may just as well create a pair of atoms with equal-and-opposite momentum, since only total momentum is conserved. So-called rogue \cite{JAV02,NAI03}, or unwanted \cite{GOR01}, dissociation to noncondensate modes therefore introduces the operators $a_{i,\pm\bk}$. 

The familiar mean-field equations are derived from the Heisenberg equation of motion for a given operator, $i\hbar\dot{x}=[x,H]$, which is derived from the Hamiltonian (not shown), and the operators are then declared $c$-numbers. In a minimalist model, $x$ represents either the $i$-th atomic amplitude $a_i$, the molecular amplitude $b$, or the so-called anomalous density, $A_\bk=a_{1,-\bk} a_{2,\bk}$, which arises from rogue dissociation to noncondensate atom pairs of equal-and-opposite momentum. The summation over $\bk$ implicit to the Hamiltonian is then converted to an integral over frequency, introducing the frequency $\omega_\rho=\hbar\rho^{2/3}/2m_r$ with $m_r=m_1m_2/(m_1+m_2)$ the reduced mass of the atom pair. All told, the mean-field equations of motion are
\bml
\bea
i\dot{a}_1 &=& -\half\Omega a_2^* b, 
\\
i\dot{a}_2 &=& -\half\Omega a_1^* b,
\\
i\dot{b} &=&  (\delta_0-i\Gamma_0/2) b
  -\half\xi\!\int\!d\varepsilon\sqrt{\varepsilon}\,f(\varepsilon)A(\varepsilon),
  \label{B_DOT}
\\
i\dot{A}(\varepsilon) &=&\varepsilon A(\varepsilon) -\Omega\,f(\varepsilon)b,
\label{A_DOT}
\eea
\label{EFF_BOSE_EQM}
\eml
where $\Omega$ is the atom-molecule coupling, $\delta_0$ is the detuning of the photoassociation laser from the $|a_1,a_2\rangle\leftrightarrow|e\rangle$ transition, and the natural linewidth of the electronically excited state is $\Gamma_0$. A microscopic fraction of molecules means that Bose enhancement of spontaneous molecular decay~\cite{GOE01} is negligible. Photodissociation is parameterized in terms of the kinetic energy of the atom pair $\hbar\varepsilon$, the coupling $4\pi\xi=\Omega/\omega_\rho^{3/2}$, and the continuum shape $f(\varepsilon)$. Whereas the standard shape for the continuum in quantum optics--say, autoionization theory~\cite{FAN61}--is a Lorentzian, here we use a Gaussian for the sake of numerical convergence, where the width is set by the semi-classical size of the molecular state.

In numerical experiments, the rate is determined from the time $\tau$ required for the atomic probability to decay to $\sim1/e$, 
\beq
R_N=\half\rho K_N=\frac{1}{\tau},
\eeq
where $K_N$ is the rate per unit density, i.e., the rate coefficient. The factor of 2 arises since we assume that losses are observed by monitoring one condensate or the other, whereas $\rho$ is the total density. Note that the lightshift correction is obtained as follows: the detuning is set to an initial nonzero guess, and the time $\tau$ required for $P_a\sim1/e$ is determined; the detuning is updated, and then $\tau$ is updated; repeating until $\tau$ is minimized.

We also develop a simple analytical result for the atom-molecule conversion rate, similar to the approach used in Refs.~\cite{MAC08,MAC10}. In short, we use the adiabatic approximation $\dot{A}=0$ first to eliminate $A$, and again in the form $\dot{b}=0$ to eliminate $b$. We thus arrive at the usual all-atom equations of motion with a tunable interaction
\bml 
\bea
\dot{a}_1&=&-\frac{\Omega^2}{4\tilde\delta}\,|a_2|^2a_1,
\\
\dot{a}_2&=&-\frac{\Omega^2}{4\tilde\delta}\,|a_1|^2a_2,
\eea
Which defines the (complex) resonant scattering length 
\beq
\rho^{-1/3}a_{\rm res}=-\frac{\Omega^2}{8\pi\omega_\rho(\delta-i\Gamma/2)}.
\eeq
Here the detuning and damping are now $\delta=\delta_0+\sigma_0$ and $\Gamma=\Gamma_0+\gamma$. The lightshift is $\sigma_0=\Re(\Sigma)$ and the photodissociation rate is $\gamma=\Im(\Sigma)$, where 
\beq
\Sigma=\lim_{\omega\rightarrow0}\left[
\quarter\Omega\xi
  \int d\varepsilon\sqrt{\varepsilon}\,\frac{f^2(\varepsilon)}{\varepsilon-\omega}.
\right]
\eeq
\eml
The rate for two-body losses from the condensate due to heteronuclear photoassociation is
\beq
R_A=\half\rho K_A=4\pi\frac{\hbar\rho}{m_r}\,\Im(a_{\rm res}).
\label{ANAL_MB}
\eeq

{\em Results.}--For concreteness, we focus on photoassociation of a joint Li-Na condensate. The atom-molecule coupling is borrowed from photoassociation of $^7$Li alone~\cite{MAC08}, so that $\Omega=\Omega_0\sqrt{\rho/\rho_0}$ with $\Omega_0=290\times2\pi$~kHz and $\rho_0=4\times10^{12}$~cm$^3$. Also from photoassociation of $^7$Li alone~\cite{MAC08}, the semi-classical size of the molecular state, which sets the width of the Gaussian $f(\varepsilon)$, is taken to be $L=133a_0$ with $a_0$ the Bohr radius, and the natural linewidth of the excited state is $\Gamma_0=12\times2\pi$~MHz.

\begin{figure}[b]
\centering
\includegraphics[width=8.5cm]{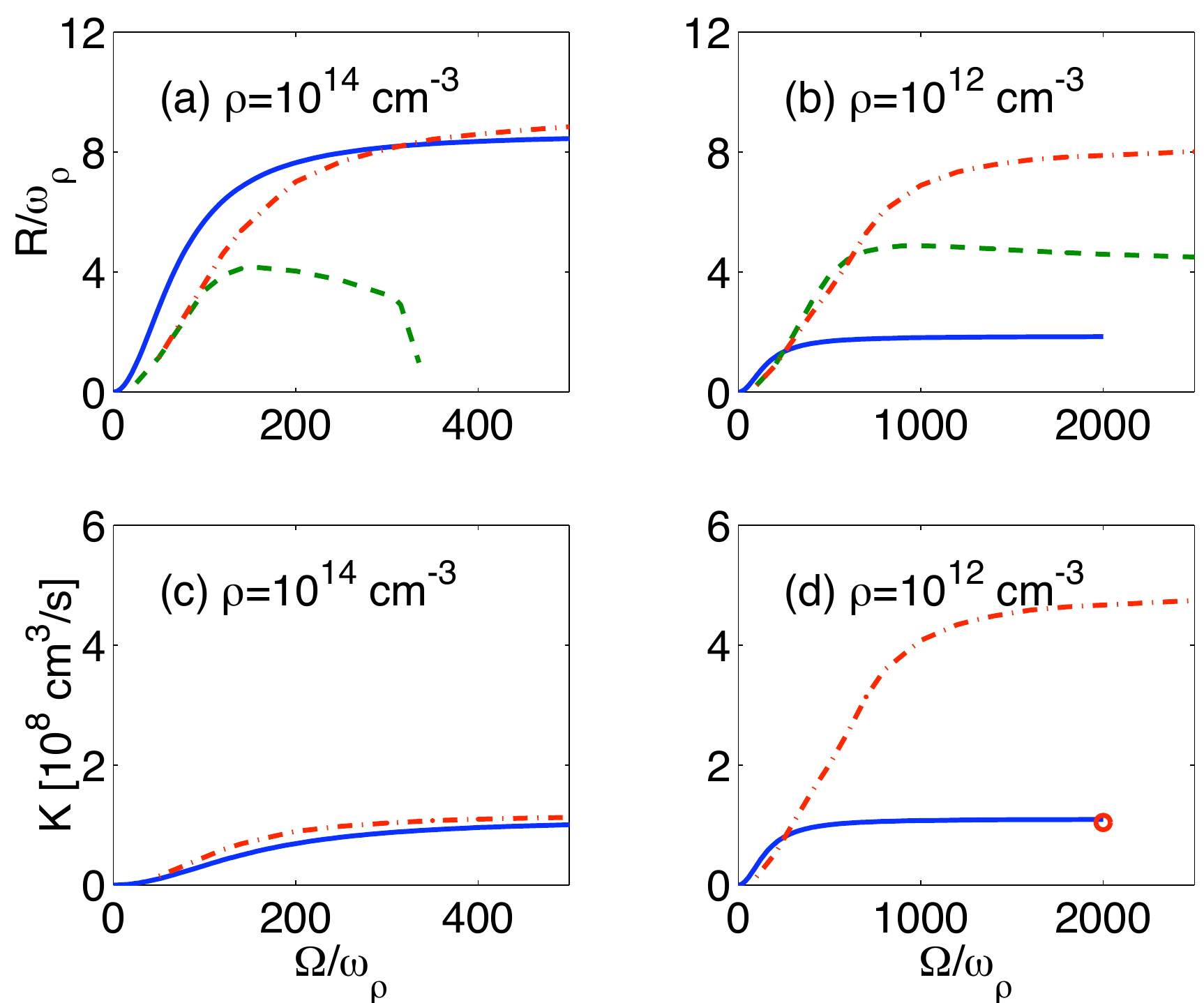}
\caption{(Color online)~Many-body rate limit for heteronuclear photoassociation of a joint Li-Na condensate. (a,b) The solid blue line is the analytical rate, and the dashed green (dot-dashed red) line is the numerical rate without (with) the unanticipated lightshift. (c,d)~Photoassociation rate constant, $K=R/\rho$, where the solid blue (dot-dashed red) line is again the analytical (lightshifted numerical) result. The open red circle in panel~(d) is given by $K_o=K_{\rm max}/(\rho_h/\rho_l)^{1/3}$, where $K_{\rm max}$ is the rate constant for $\rho_l=10^{12}$~cm$^3$ and $\rho_h=10^{14}$~cm$^3$.}
\label{RATE_LIM}
\end{figure}

Results are shown in Fig.~\ref{RATE_LIM} for $\rho=10^{14}$~cm$^{-3}$ [panel~(a,c)] and $10^{12}$~cm$^{-3}$ [panel~(b,d)]. For lasers tuned to the usual lightshifted resonance~\cite{FED96}, $\delta_0=-\sigma_0$, the numerical rate maximizes at a critical atom-molecule coupling $\Omega_c$ (Fig.~\ref{RATE_LIM}, dashed green lines), at a value that increases with decreasing density, $R_{\rm max}\sim4\omega_\rho$ for $\rho=10^{14}$~cm$^{-3}$ and $R_{\rm max}\sim5\omega_\rho$ for $\rho=10^{12}$~cm$^{-3}$. This density dependence is similar to Ref.~\cite{JAV02}, which includes the lightshift $\sigma_0$. Correcting the lightshift according to $\sigma=\sigma_0+1.2(\Omega/\Omega_c)^2\Gamma_0$ leads to saturation instead of maximization (Fig.~\ref{RATE_LIM}, dot-dashed red lines). The precipitous drop in the numerical rate in panel~(a) arises because, absent the lightshift, the strongly-coupled system undergoes coherent oscillations between atomic condensate and photodissociated pairs, and the atomic probability therefore takes longer to drop to $P_a\sim1/e$. Such oscillations are absent (on the timescale $\tau$) once the lightshift is corrected. Moreover, the corrected numerical rate saturates at a value that increases slightly for increasing density, $R_{\rm max}/\omega_\rho=8.2$~(8.8) for $\rho=10^{14(12)}$~cm$^{-3}$, in contrast to the unshifted result and Ref.~\cite{JAV02}. The rate constant scales with density as $K\propto\rho^{-1/3}$, as indicated by the open red circle in Fig.~\ref{RATE_LIM}(b). Also, the analytical results (solid blue lines) agree best with the numerical results for high density. Finally, $R/\omega_\rho$ depends very weakly on density, so that inhomogeneous effects, which we have neglected, should be marginal.

\begin{figure}[b]
\centering
\includegraphics[width=4.25cm]{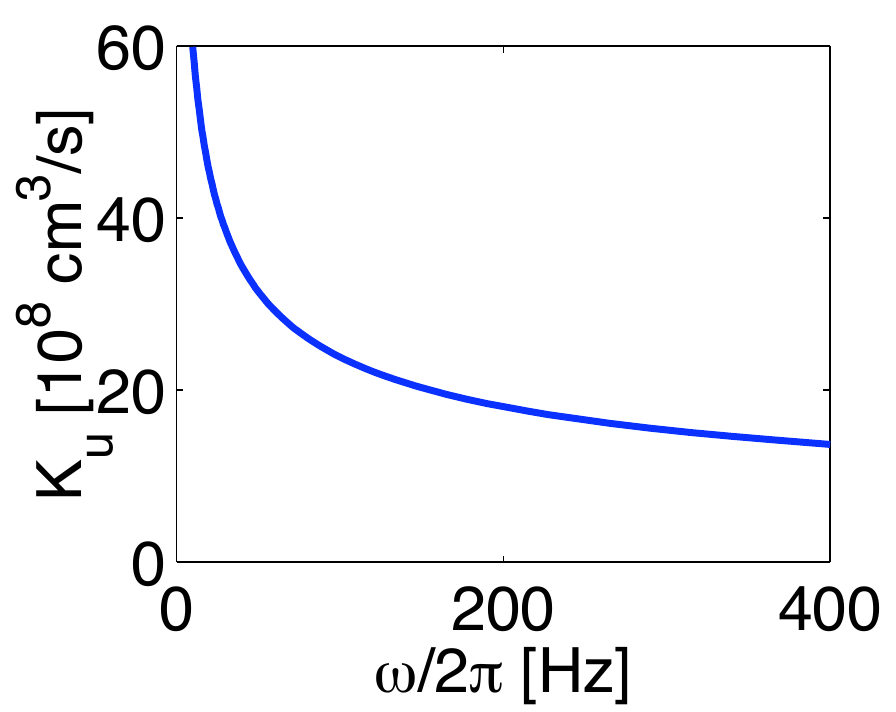}
\caption{Unitary-limited photoassociation rate constant for a Bose-Einstein condensate vs. trap frequency, as per the text.}
\label{U_RATE_LIM}
\end{figure}

We also compare the many-body rate to the two-body unitary limit, which is set by the de~Broglie wavelength of the condensate atoms according to
\beq
R_u/\rho=K_u=\frac{\hbar}{m_r}\,\Lambda_D.
\eeq
In the zero-temperature limit, the de~Broglie wavelength is given by the mean Thomas-Fermi radius of the condensate
\bml
\beq
\Lambda_D=2R_{\rm TF}=\sqrt{\frac{8\hbar\mu}{m\bar\omega^2}},
\eeq
where $\bar\omega^3=\omega_x\omega_y\omega_z$ defines the mean trapping frequency, and we assume that the frequencies can be adjusted so that the two condensates have identical sizes. The condensate chemical potential is
\beq
\mu=\half\bar\omega\,\left(\frac{15Na}{L_{\rm ho}}\right)^{(2/5)},
\eeq
\eml
where $N$ is the number of atoms in the condensate, $a$ is the s-wave scattering length~\cite{A_PA}, and $L_{\rm ho}=\sqrt{\hbar/(m\bar\omega)}$ is the harmonic oscillator length scale. Results are shown in Fig.~\ref{U_RATE_LIM} for trap frequencies satisfying $\sqrt{2}\omega_x=\omega_y=\omega_z/\sqrt{2}$ and $N_i=10^6$ atoms. For loose traps, i.e., low density, the two-body unitary limit set by the condensate size is over an order of magnitude larger than the many-body rate limit.

{\em Conclusion}--We have investigated the rate limit on photoassociation of a Bose-Einstein condensate, using an upgraded many-body model to obtain both numerical and analytical results. An unanticipated lightshift leads to a maximum in the numerical photoassociation loss rate for strong atom-molecule coupling, similar to the many-body model without spontaneous decay~\cite{JAV02}. Once this lightshift is taken into account, the numerical rate saturates at about $9(\hbar\rho^{2/3}/2m_r)$, so that the many-body rate constant for resonant photoassociation scales with density as $K\propto\rho^{-1/3}$. This limit holds for over two decades in densities, whereas recent analytical results~\cite{NAI03} predict this density dependence for limited densities. It also agrees with the rate limit obtained for combined photoassociation and Feshbach resonances~\cite{MAC10}. Our own analytical approximation yields a rate constant that is independent of density, and agrees best with the numerical result for dense condensates. Finally, the many-body rate limit has been shown explicitly to be generally more strict than a two-body unitary rate limit.

\acknowledgements{Supported by the National Science Foundation}


\begin{thebibliography}{99}
\bibitem{A_PA}
Although photoassociation is known to change the s-wave scattering length, condensate expansion due to a changing scattering length occurs on the timescale of the motion of the atoms in the trap, which is much longer than the timescale for photoassociation, and we therefore take the s-wave scattering length as the background (zero-field) value.
\end{thebibliography}
\end{document}